\documentclass{article}

\usepackage{arxiv}

\usepackage[utf8]{inputenc} 
\usepackage[T1]{fontenc}    
\usepackage{hyperref}       
\usepackage{url}            
\usepackage{booktabs}       
\usepackage{amsfonts}       
\usepackage{nicefrac}       
\usepackage{microtype}      
\usepackage{lipsum}		
\usepackage{graphicx}
\usepackage[square,numbers]{natbib}
\usepackage{doi}

\usepackage[export]{adjustbox}
\usepackage{tablefootnote}
\usepackage{xcolor}
\usepackage{graphicx}
\usepackage{cuted}
\usepackage{float}
\usepackage{graphicx}
\usepackage{caption}
\usepackage{comment}
\usepackage{amsmath}
\definecolor{color}{RGB}{25,25,112}
\definecolor{negro}{RGB}{0,0,0}
\definecolor{colorurl}{RGB}{25,25,112}
\usepackage{orcidlink}
\usepackage{amssymb}
\usepackage{tikz}
\usepackage{tikz-3dplot}
\usetikzlibrary{patterns}

\usepackage{ulem}

\usepackage{chemformula} 

\usepackage[T1]{fontenc}

\title{Framework for extracting the rates of photophysical processes from biexponentially decaying photon emission data}

\author{\href{https://orcid.org/0000-0001-8047-5660}{\includegraphics[scale=0.06]{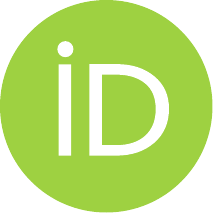}\hspace{1mm} Jill M. Cleveland}\href{https://orcid.org/0000-0002-9008-9322}{\includegraphics[scale=0.06]{orcid.pdf}\hspace{1mm} Tory A. Welsch}
\href{https://orcid.org/0000-0000-0000-0000}{\includegraphics[scale=0.06]{orcid.pdf}\hspace{1mm} Eric Y. Chen} \href{https://orcid.org/0000-0000-0000-0000}{\includegraphics[scale=0.06]{orcid.pdf}\hspace{1mm} D. Bruce Chase} \href{https://orcid.org/0000-0001-7999-3567}{\includegraphics[scale=0.06]{orcid.pdf}\hspace{1mm} Matthew F. Doty} \thanks{doty@udel.edu} \\
	Department of Materials Science and Engineering, \\ University of Delaware, Newark, DE 19716, USA \AND \href{https://orcid.org/0000-0003-1575-6583}{\includegraphics[scale=0.06]{orcid.pdf}\hspace{1mm} Hanz Y. Ram\'irez-G\'omez} \thanks{hanz.ramirez@uptc.edu.co} \\
	Grupo de F\'isica Te\'orica y Computacional \& Grupo QUCIT, \\
	Escuela de F\'isica, Universidad Pedag\'ogica y Tecnol\'ogica de Colombia (UPTC),\\
	Tunja 150003, Boyac\'a, Colombia. \\
}

\date{}

\renewcommand{\shorttitle}{\textit{arXiv} Template}

\begin{document}
	\maketitle
	
\begin{abstract}
There is strong interest in designing and realizing optically-active semiconductor nanostructures of greater complexity for applications in fields ranging from biomedical engineering to quantum computing. While these increasingly complex nanostructures can implement progressively sophisticated optical functions, the presence of more material constituents and interfaces also leads to increasingly complex exciton dynamics. In particular, the rates of carrier trapping and detrapping in complex heterostructures are critically important for advanced optical functionality, but they can rarely be directly measured. In this work, we develop a model that includes trapping and release of carriers by optically inactive states. The model explains the widely observed biexponential decay of the photoluminescence signal from neutral excitons in low dimensional semiconductor emitters. The model also allows determination of likelihood intervals for all the transition rates involved in the emission dynamics, without the use of approximations. Furthermore, in cases for which the high temperature limit is suitable, the model leads to specific values of such rates, outperforming reduced models previously used to estimate those quantities. We demonstrate the value of this model by applying it to time resolved photoluminescence measurements of CdSeTe/CdS heterostructures. We obtain values not only for the radiative and nonradiative lifetimes, but also for the delayed photoluminescence originating in trapping and release. 
\end{abstract}


\section{Introduction}\label{Sect:Intro}
Radiative recombination between electrons and holes localized in the discrete conduction and valence band states of low dimensional semiconductor nanostructures is well established and has been extensively studied. A particular advantage of semiconductor nanostructures is that their structure, size, and composition can be adjusted to tailor the optical absorption, carrier transfer, and optical emission for applications ranging from fundamental research on quantum information processing to state-of-the-art medical treatments \cite{intro1,intro2,intro3,intro4,intro5,intro5.5,intro6}. In particular, there is strong present interest in using increasingly complex heterostructures to control band offsets, band alignments, and carrier transfer in order to realizing increasingly complex optical functionality (e.g. references ~\cite{Albers2012,complex-0,ACSNANO-2018,complex-1,complex-2,complex-3,complex-4,complex-5}). Optimizing such structures for specific optoelectronic applications requires a detailed knowledge of the dynamic processes behind photon emission. 

One of the most common methods used to study carrier dynamics in semiconductor nanostructures is to measure the time dependence of the photoluminescence (PL) signal, which is expected to be proportional to the population of the optically-excited state. In the ideal case of 100 \% quantum yield, the population decay will be related only to the radiative recombination rate and will be monoexponential. Such an ideal scenario is rarely achieved and generally requires cryogenic temperatures \cite{idealyield1,thermal-effects,idealyield2}. Nonradiative decay channels typically reduce the luminescence efficiency and increase the decay rate of the time resolved photoluminescence signal (TRPL). When a fast nonradiative decay channel is present, the measured exciton lifetime decreases, but the TRPL decay still has single exponential form (i.e. $\tau_{PL} = 1/\Gamma_{PL}$, $\Gamma_{PL} = \Gamma_{radiative} + \Gamma_{nonradiative}$). In this case it is not possible to obtain distinct values for the radiative and nonradiative decay rates from TRPL experiments alone, but separate rates can be obtained if accurate measurements of the photoluminescence quantum yield (PLQY) $\Phi$ are also available \cite{average-0,average-1}.      

The issue of separating nonradiative and radiative decay rates becomes more complex and interesting when multiexponential decay is observed in the TRPL spectra \cite{multiexp-1,multiexp-2}. For example, biexponentially-decaying TRPL is regularly reported for different types of nanostructures including colloidal II-VI nanoparticles \cite{biexp-II-VI-1,biexp-II-VI-2}, quantum wells \cite{biexp-qw-1,biexp-qw-2}, metal-halide perovskites \cite{biexp-perovskites-1,biexp-perovskites-2}, self-assembled III-V quantum dots \cite{biexp-self-assembled-1,biexp-self-assembled-2}, organic nanoparticles \cite{biexp-organic1,biexp-organic2}, lithographically defined nanostructures \cite{biexp-lithography-1,biexp-lithography-2},  and monolayer semiconductors \cite{biexp-monolayer-1,biexp-monolayer-2}. There are two main explanations for such a biexponential decay: i) the PL signal is the superposition of two independent radiative process, each of which has a distinct monoexponential radiate decay rate \cite{biexp-organic2}, or ii) there is some physical process that introduces an additional time scale in the emission from the radiative channel of interest, including the presence of delayed emission from carriers that are captured by and later released from traps \cite{multiexp-1}. These two explanations are not exclusive and both situations could coexist, particularly in ensemble measurements \cite{biexp-II-VI-2}. If situation (i) is dominant, the measurement of the amplitude coefficients of each decay process in a multi-exponential fit can be used to establish the fraction of photons emitted via each radiative channel. If situation (ii) is dominant, the trapping times are contributing to the experimentally measured bi-exponential decay times but it is challenging to extract, from the data, separate values for the rates of trapping, radiative decay, and nonradiative decay.

In this paper we develop a model that explains the biexponential decay of the photoemission from luminescent nanostructures, in terms of four participating energy levels:  the ground state, a state representing the single radiative channel, a state representing the main delaying trap state and a state accounting for nonradiative energy dissipation. This model allows analysis of the dynamics within the system by utilizing PLQY and TRPL data to extract rates for both radiative and nonradiative processes. It is pertinent for light emitters exhibiting a biexponential time-dependent PL, in which there is a dominant emitting state, and in which carrier trapping by a representative optically-inactive state is expected to play a significant role on the population of the radiative state (situation ii)). We also apply the developed model on some suitable heterostructures. In section \ref{Sect:model} we develop and analytically solve a system of coupled equations describing all of the possible transitions between the discrete states of a single nanoparticle without making the simplifying approximations that have been made in prior reports of similar models. We show that our model is both consistent with and transcends the prior relevant models because it is applicable to a wider range of systems and conditions. In section \ref{Sect: Analysis} we show how the solutions of this model enable the extraction of the rates of underlying physical processes from data obtained via TRPL and quantum yield experiments. In particular, we show that  the model provides likelihood intervals for the values of the relevant transition rates that underlie the system dynamics and that under some circumstances specific values can be extracted. In section \ref{Sect:Experimental} we introduce an experimental test system consisting of two samples, each of which is an ensemble of colloidally-synthesized semiconductor heterostructures with specific differences between the samples that were designed to influence the rates of carrier transfer within the nanostructure. In section \ref{Sect:application} we demonstrate the value of our model and analysis framework by applying it to experimental TRPL and quantum yield data obtained from these samples. In section \ref{Sect:Conclusion} we summarize the results. The most important result is the creation of a clear framework for the extraction of trapping, radiative, and nonradiative decay rates from multi-exponential TRPL decay data that can be applied to a wide range of materials. 

\section{Theoretical model}\label{Sect:model}

\subsection{Involved States and the Resonant Excitation Limit}

\begin{figure}[h]
	\includegraphics[scale=0.45]{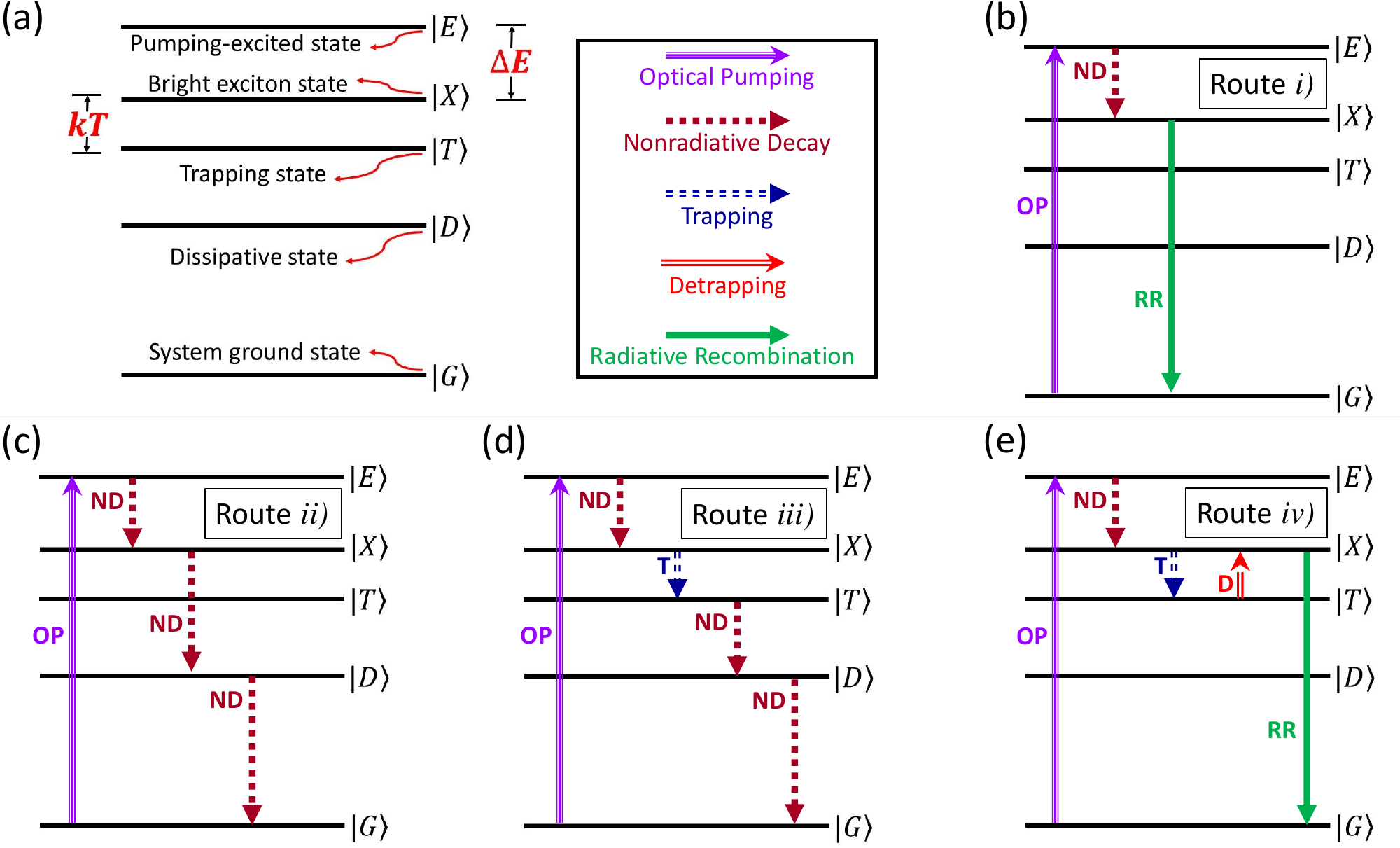}
	\caption{(a) States involved in the photoluminescence processes we model. The legend explains what type of transition is indicated by each arrow type. (b) - (e) Four possible routes by which carriers move from the excited to the ground state. The pathways in panels (b) and (e) result in photon emission while the ones depicted in panels (c) and (d) are entirely nonradiative.}
	\label{figure-1}
\end{figure}

In order to explain the observed biexponential decay of a PL signal, we start our modeling by assuming that the photoluminescence process involves five states, as shown in figure \ref{figure-1}(a). These states are: 1) the ground state $ | G \rangle $, which represents the carriers within the single nanostructure in their lowest energy configuration. 2) The excited state $| E \rangle $ that is populated by photoexcitation. 3) The bright exciton state $ | X \rangle $, which is below $| E \rangle $ in energy by an amount $ \Delta E $ that depends on the wavelength of the exciting light. 4) The trapping state $ | T \rangle $, which represents states with a very small dipole matrix element such that they effectively trap carriers with negligible probability of radiative recombination. The energy difference between state $ | T \rangle $ and state $ | X \rangle $ is smaller than the thermal energy $k_B T $, where $k_B$ is the Boltzmann constant and $T$ is the temperature of the system, which means that thermally-driven transitions between those states are possible. This energy difference can be either positive or negative. 5) The dissipative state $ | D \rangle$ from which there is also no photon emission, but from which thermal escape back to the $ | X \rangle $ state is not possible. The transfer $ | D \rangle \rightarrow  | X \rangle $ may be forbidden either because the energy eigenvalues of $ | D \rangle$ are lower than those of $ | X \rangle $ by more than $k_B T $ or by angular momentum conservation. In all cases, carriers can leave the $ | D \rangle$ state only by nonradiative relaxation back to $ | G \rangle $.   

For the model of available states depicted in figure \ref{figure-1} there are two pathways that result in radiative recombination and emission of photons. Route \textit{i)} (panel (b)), involves the direct transition $ | X \rangle \rightarrow  | G \rangle $ that occurs after $ | X \rangle $ is populated by the excitation laser. Route \textit{iv)} (panel (e)), results in similar photon emission as route \textit{i)}, but has a longer recombination time because of the delay related to trapping and detrapping (i.e. $ | X \rangle \rightarrow  | T \rangle $ is followed by $ | T \rangle \rightarrow  | X \rangle $ and then $ | X \rangle \rightarrow  | G \rangle $). There are also two pathways that result in nonradiative decay. Route \textit{ii)} (panel (c)) describes the process in which, after pumping, carriers relax nonradiatively via $ | X \rangle \rightarrow  | D \rangle $ and then $ | D \rangle \rightarrow  | G \rangle $). Route \textit{iii)} (panel (d)) similarly involves the dissipative state and does not end in photon emission, but decays via the trapping state $|T\rangle$ (i.e. $ | X \rangle \rightarrow  | T \rangle $, followed by $ | T \rangle \rightarrow  | D \rangle $ and then $ | D \rangle \rightarrow  | G \rangle $). A TRPL experiment detects the photons emitted by processes \textit{i)} and \textit{iv)}, but the rates and probabilities of processes ii) and iii) influence the measured decay times. 

It is known that resonant excitation of the $|X\rangle$ state substantially increases the PLQY \cite{resonance1,resonance2,TRPL-2021}. This is essentially due to suppression of the nonradiative losses that can happen during the decay $|E\rangle \rightarrow |X\rangle$. In this work we consider an ideal scenario where the pumping resonantly excites the exciton state ($|E\rangle \equiv |X\rangle$ and $\Delta E \rightarrow 0$). In other words, we assume that there is always rapid nonradiative relaxation from state $| E \rangle $ to state $ | X \rangle $ on timescales faster than any other rate in the system of equations and with 100\% probability. While such a simplification neglects one possible set of dissipative pathways, it also keeps the model analytically solvable and allows for quantitative analysis of the relationship between the measurable decay times and the transition rates between the states $|X\rangle$, $|T\rangle$, $|D\rangle$ and $|G\rangle$. Through the remainder of the work presented here we thus neglect state $| E \rangle $ and consider only the four states $|X\rangle$, $|T\rangle$, $|D\rangle$ and $|G\rangle$.

\subsection{System of equations}

\begin{figure}[H]
	\includegraphics[scale=0.3]{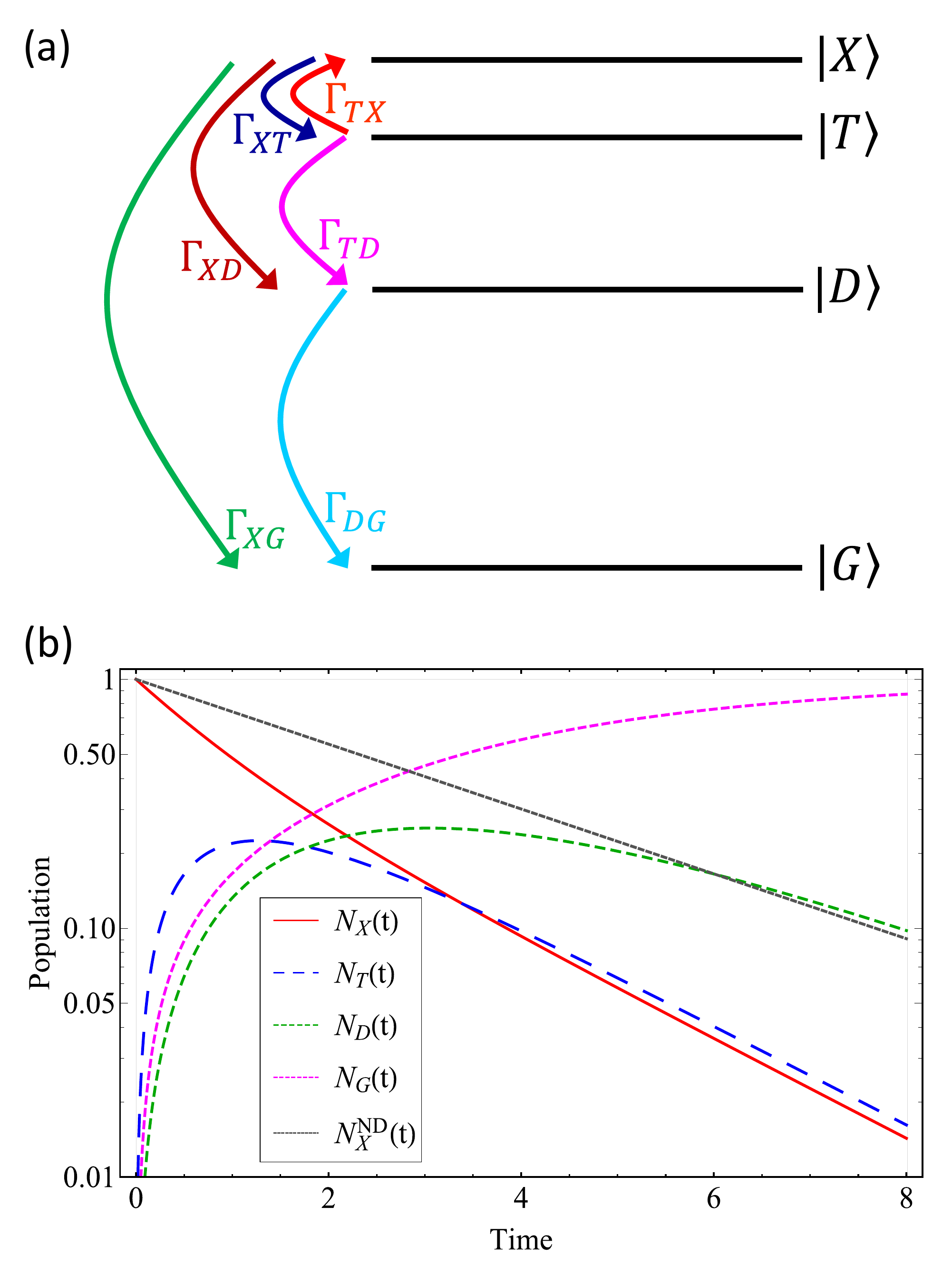}
	\caption{(a) Transition rates involved in the modeled dynamics. (b) Time dependent population of the considered states for arbitrary values of the rates ($\Gamma_{XT}=0.5$, $\Gamma_{XD}=0.1$, $\Gamma_{XG}=0.2$, $\Gamma_{TX}=0.3$, $\Gamma_{TD}=0.6$ and $\Gamma_{DG}=0.4$). To highlight the contrast with the delayed (biexponential) exciton population (red-solid line), the corresponding monoexponential non-delayed population  $N_X^{ND}(t)$, i.e. the limit in which  $\Gamma_{XT}=\Gamma_{XT}=\Gamma_{TD}=0$,  is also shown (black-dashed line). }
	\label{figure-2}
\end{figure}

To describe the time dependent emission of photons by exciton recombination, we assume that the number of emitted photons is proportional to the instantaneous population of state $| X \rangle$ \cite{proportionaltopopulation}. Thus, a time dependent expression for the population of the $| X \rangle$ state can be used to correlate experimental measurements with model parameters such as the involved transition rates. To find such an expression, we solve the system of coupled equations

\begin{equation}
	\dot{\boldsymbol{N}}(t) = \tilde{\Gamma} \boldsymbol{N}(t)  \hspace{1ex}, 
	\label{systemofequations}
\end{equation}

where the population vector is defined as 

\begin{equation}
	\label{populations}
	\boldsymbol{N}(t)  =  \left(
	\begin{array}{c}
		N_X (t)  \\
		N_T (t) \\
		N_D (t) \\
		N_G (t)
	\end{array}
	\right) \hspace*{1ex} ,
\end{equation} 

and the rate matrix is given by

\begin{equation}
	\label{rate-matrix}
	\tilde{\Gamma}  =  \left(
	\begin{array}{cccc}
		- \Gamma_{XT} - \Gamma_{XD} - \Gamma_{XG} &  \Gamma_{TX}  & 0  & 0 \\
		\Gamma_{XT}  & - \Gamma_{TX} - \Gamma_{TD}   & 0  & 0 \\
		\Gamma_{XD}   & \Gamma_{TD}  & - \Gamma_{DG}  & 0 \\
		\Gamma_{XG} & 0  & \Gamma_{DG}  & 0
	\end{array}
	\right) \hspace*{1ex} ,
\end{equation} 

in terms of the relevant transition rates, which are shown in figure \ref{figure-2}(a).

The general solution of equation (\ref{systemofequations}) has the form

\begin{equation}
	\label{general}
	 \boldsymbol{N}(t)  = \sum_{i=1}^{4} \alpha_i e^{\lambda_i t} \boldsymbol{V}_i  	\hspace*{1ex} ,
\end{equation} 

where $\lambda_i$ and $\boldsymbol{V}_i$ are respectively the eigenvalues and eigenvectors of matrix $\hat\Gamma$, and $\alpha_i$ are functions that depend on the transition rates and on the initial conditions.  

Because the matrix in equation (\ref{rate-matrix}) does not have any particular symmetry, there are not simple expressions for the functions $\alpha_i$. Prior work has reported solutions only under certain limiting assumptions \cite{PhysRevLett.high-T,temperature-dependence,added-4x4-model,multiexp-2}. Here we solve this system of equations analytically without making any of these limiting approximations. Relevant details of our solution are provided in sections 1 and 2 of the Supporting Information \cite{supp-mat}. In the remainder of this section we summarize the results and describe how they go beyond prior solutions made under limiting assumptions. 

Under the initial conditions $N_X(0)=1$ and $N_T(0)=N_D(0)=N_G(0)=0$, after lengthy algebra, diagonalization of the matrix in equation (\ref{rate-matrix}), yields the normalized population functions

\begin{eqnarray}
	N_X(t) &=& A^{-}_X e^{- \Gamma_- t} + A^{+}_X e^{- \Gamma_+ t}  \hspace*{1ex} , \nonumber \\
	N_T(t) &=& A_T \left( e^{- \Gamma_- t} - e^{- \Gamma_+ t} \right) \hspace*{1ex} , \nonumber \\
	N_D(t) &=& A^{-}_D e^{- \Gamma_- t} + A^{+}_D e^{- \Gamma_+ t} + A^{0}_D e^{- \Gamma_{DG} t} \hspace*{1ex} \nonumber \\
	N_G(t) &=& A^{-}_G e^{- \Gamma_- t} + A^{+}_G e^{- \Gamma_+ t} + A^{0}_G e^{- \Gamma_{DG} t} + 1 \hspace*{1ex} , 
	\label{solutions}
\end{eqnarray}

written in terms of the composed rates $\Gamma_+$ and $\Gamma_-$,  that are defined according to  

\begin{eqnarray}
	\Gamma_+ &=& \frac{\Gamma_1 + \Gamma_2}{2} \hspace*{1ex} , \nonumber \\
	\Gamma_- &=& \frac{\Gamma_1 - \Gamma_2}{2} \hspace*{1ex} , 
	\label{rates-1}
\end{eqnarray}

where

\begin{eqnarray}
	\Gamma_{*} &=& \Gamma_{XG} + \Gamma_{XD} \hspace*{1ex} , \nonumber \\
	\Gamma_{0}^2 &=& \Gamma_{*} \Gamma_{TX} + \Gamma_{*} \Gamma_{TD} + \Gamma_{XT} \Gamma_{TD} \hspace*{1ex} , \nonumber \\
	\Gamma_1 &=& \Gamma_{*} + \Gamma_{XT} + \Gamma_{TX} + \Gamma_{TD} \hspace*{1ex} , \nonumber \\
	\Gamma_2 &=& \sqrt{\Gamma_1^2 - 4 \Gamma_0^2} \hspace*{1ex} .
	\label{rates-2}
\end{eqnarray}

The decay rates $\Gamma_+$ and $\Gamma_-$ are to be associated to the corresponding time constants that would be obtained from biexponential fits to measured TRPL from suitable samples. It is important to highlight that expressing them in terms of $\Gamma_{*}$, $\Gamma_{0}$, $\Gamma_{1}$ and $\Gamma_{2}$, which are compact functions of the involved rates, is a useful accomplishment because it allows to relate experimentally obtained data with the parameters of the dynamical model.      

For any positive values of the rates $\Gamma_{*}$, $\Gamma_{XT}$, $\Gamma_{TX}$, and $\Gamma_{TD}$, the inequalities $\Gamma_1^2 > 4 \Gamma_0^2 $ and $\Gamma_{1} > \Gamma_{2}$ hold, so that $\Gamma_{2}$ is always real and $\Gamma_{-}$ is definitely positive. 

The amplitude coefficients for the decay of the population of state $| X \rangle$ are given by

\begin{eqnarray}
	A^{-}_X &=& \frac{1}{2} - \frac{ \Gamma_{*} + \Gamma_{XT}  - \left( \Gamma_{TX} + \Gamma_{TD} \right)}{2 \Gamma_{2}}  \hspace*{1ex} , \nonumber \\
	A^{+}_X &=& \frac{1}{2} + \frac{\Gamma_{*} + \Gamma_{XT} - \left( \Gamma_{TX} + \Gamma_{TD} \right)}{2 \Gamma_{2}}  \hspace*{1ex} . 
	\label{amplitudes-X}
\end{eqnarray}

It is worth noting that either equation $A^{-}_X$ or $A^{+}_X$, but not both, adds actual information. This because the required normalization imposes $A^{-}_X + A^{+}_X =1$. Hence, having an expression for both $A^{-}_X$ or $A^{+}_X$ helps only to check the consistency of the solution. For example, it can be verified by straightforward substitution that they indeed satisfy $N_X(0)=A^{-}_X + A^{+}_X=1$. 

The amplitude of the other population functions that appear in equation (\ref{solutions}) are algebraically more complicated and not essential for the rest of our discussion here. All of these amplitude functions are presented in subsection 1.1 of the Supporting Information \cite{supp-mat}, and their explicit expressions may be useful in understanding the results of experiments different from those used here to illustrate the application of this model. For instance, excitation-power dependent studies might access the population of the state $| T \rangle$ with respect to that of the state $| X \rangle$ \cite{trap-filling-1,trap-filling-2,trap-filling-3}, which could be analyzed using the expression obtained from $N_T(t)/N_X(t)$. One thing to notice is that while $N_D(t)$ and $N_G(t)$ depend explicitly on $\Gamma_{DG}$, $N_X(t)$ and $N_T(t)$ are completely insensitive to the transition rate between $| D \rangle $ and $| G \rangle $. 

We stress that equations (\ref{solutions}) - (\ref{amplitudes-X}) are analytical solutions that do not include any approximations of the involved rates. Equations (\ref{solutions}) have elements in common with those reported previously, but the previous models are less general because they are built under certain limiting suppositions \cite{PhysRevLett.high-T, temperature-dependence, multiexp-2, added-4x4-model}. For example, Labeau et al. assume $\frac{\Gamma_{XT} \Gamma_{TX}}{\Gamma_{XG} \Gamma_{TD}} \gg 1$ \cite{PhysRevLett.high-T}. In contrast, Gokus et al. assume $\frac{\Gamma_{XT} \Gamma_{TX}}{\Gamma_{XG} \Gamma_{TD}} \ll 1$ \cite{temperature-dependence}. Finally, Gong et al. neglects $\Gamma_{TD}$ from the beginning  \cite{multiexp-2}, while Murphy et al. do the same not only with $\Gamma_{TD}$ but also with $\Gamma_{TX}$ \cite{added-4x4-model}. Those suppositions are equivalent to asymptotic limits within which, to obtain an analytical solution for the system of coupled equations is easier, and for which the final expressions are plainer. However, such simplifications imply inability to explore intermediate regimes and then, reduced generality. Thus, the model presented here is both consistent with and transcends the prior relevant models because it is applicable to a wider range of systems and conditions. 

According to equations (\ref{solutions}) the solution for $N_X (t)$ is a biexponential function with two characteristic rates ($\Gamma_+$ and $\Gamma_-$). These rates correspond to the ``short'' and ``long'' time constants obtained from a biexponential fit to TRPL data. Which component of the biexponential population has a greater amplitude depends on the sign of the quantity $\delta \equiv (\Gamma_{*} + \Gamma_{XT}) - (\Gamma_{TX} + \Gamma_{TD})$. This biexponential behavior fits many experimental reports on different types of semiconductor nanostructures (e.g.~\cite{GaN-nanowires,PbS-QDs,ZnAg-QDs,orange-thin-films,CdTe-nanocrystals,ZnCdTeS-nanocrystals}), including the measurements of our test structures as described below. 

For the sake of visualization, figure \ref{figure-2}(b) shows the time dependence of the populations $N_X (t)$, $N_T (t)$, $N_D (t)$ and $N_G (t)$ for some arbitrary dimensionless values $\Gamma_{XT}=0.5$, $\Gamma_{XD}=0.1$, $\Gamma_{XG}=0.2$, $\Gamma_{TX}=0.3$, $\Gamma_{TD}=0.6$ and $\Gamma_{DG}=0.4$. Those particular numbers are chosen to help distinguishing the corresponding populations and to highlight the contrast between the conventional and the improved models. The key point is the observation that the measured TRPL would be proportional to $N_X (t)$, which has a biexponential decay as a result of the complex carrier transfer dynamics introduced by the presence of the trapping state. For comparison, if the optically inactive trapping state is removed from the system ($\Gamma_{XT} = \Gamma_{TX} = \Gamma_{TD} = 0$), the ``nondelayed (ND)'' population of $N_X (t)$, as shown by the black dashed line in figure \ref{figure-2}(b), recovers a monoexponential decay and the population of the exciton state would be given by (see subsection 1.2 of the Supporting Information \cite{supp-mat})

\begin{equation}
	\begin{split}
		N_X^{ND} (t) =& e^{- \Gamma_{PL} t } \hspace{1ex}, 
	\end{split}
	\label{NXnd}
\end{equation}

where $\Gamma_{PL} \equiv \Gamma_{*} = \Gamma_{XG} + \Gamma_{XD}$ is the PL decay rate ($\tau_{PL} = 1/\Gamma_{PL}$) \cite{recombination-QDs,single-GaN-nanowires,single-layer-MoS2,InGaN-dots-2020,upconverting-dots-2021}. In other words, 1) our model allows us to understand clearly how the presence of a trapping state creates biexponential decay dynamics and 2) our model reduces to the well-known monoexponential behavior in the absence of the trapping state.

Finally, given the biexponential decay of the exciton population shown by equation (\ref{solutions}), the corresponding quantum yield can be written in the form (see subsection 1.3 of the Supporting Information for details \cite{supp-mat}) 

\begin{equation}
	\begin{split}
		\Phi = \frac{ \left( \Gamma_{-} A_X^+ + \Gamma_{+} A_X^- \right) \Gamma_{XG}}{ \Gamma_{+} \Gamma_{-} } = \frac{\left( \Gamma_{TX} + \Gamma_{TD} \right) \Gamma_{XG}}{ \Gamma_{TX} \Gamma_{PL} + \Gamma_{TD} \Gamma_{PL} + \Gamma_{TD} \Gamma_{XT} } \hspace{1ex}. 
		\label{delayed-yield-fvv}
	\end{split}
\end{equation}

Importantly, this equation allows us to obtain $\Gamma_{XG}$ in terms of the quantities $\Phi$, $\Gamma_-$, $\Gamma_+$, $A_X^-$ and $A_X^+$, all of which can be obtained from experimental data for quantum yield and TRPL decay. Specifically, $\Phi$ is the measured quantum yield and $\Gamma_-$, $\Gamma_+$, $A_X^-$ and $A_X^+$ are the amplitude coefficients and time constants obtained from a biexponential fit to TRPL data. 

\section{Analysis of the obtained equations}\label{Sect: Analysis}
We would now like to invert this system of equations in order to be able to directly calculate the rates of underlying physical process from the quantities extracted from experimental measurements. The ideal situation would be to express the rates $\Gamma_{XT}$, $\Gamma_{XD}$, $\Gamma_{TX}$ and $\Gamma_{TD}$ in terms of the experimentally accessible quantities $\Phi$, $\Gamma_-$, $\Gamma_+$, $A_X^-$ and $A_X^+$. To this end, rearrangement of equation (\ref{delayed-yield-fvv}) yields a unique expression for $\Gamma_{XG}$ in the form 

\begin{equation}
	\Gamma_{XG} =  \frac{ \Phi \Gamma_{+} \Gamma_{-} }{ \Gamma_{-} A_X^+ + \Gamma_{+} A_X^-} \hspace{1ex}. 
	\label{cleared-Gxg}
\end{equation}

However, inverting the full system of equations to obtain values for the four parameters $\Gamma_{XT}$, $\Gamma_{XD}$, $\Gamma_{TX}$ and $\Gamma_{TD}$ is not possible because there are only three independent equations: those for $\Gamma_{-}$, $\Gamma_{+}$, and \textit{either} $A_X^-$ \textit{or} $A_X^+$, which are not linearly independent because $A^{-}_X + A^{+}_X=1$. We note that the system of equations also includes a fifth parameter $\Gamma_{DG}$, but this becomes irrelevant in studying the PL dynamics because the population $N_X(t)$ is independent of it. Physically this is because carriers lost to the dissipative state $| D \rangle$ will never return to $| X \rangle$ no matter how long they take to decay back to $| G \rangle$.

In what follows we show how the obtained relations provide physically meaningful information in the form of constraints on the possible values for the considered rates even when complete inversion of the system of equations is impossible. We also discuss how specific values can be determined for the involved rates under specific conditions that allow for the approximation $\Gamma_{XT} \sim \Gamma_{TX}$. 

\subsection{Expressions in terms of experimentally-accessible quantities}

After some algebra detailed in section 2 of the Supporting Information \cite{supp-mat}, we can rewrite the available equations as 

\begin{equation}
	\begin{split}
		\Gamma_{TX} =& \frac{\Gamma_{C}^2 - \Gamma_{4} \Gamma_{TD}}{\Gamma_{PL}}   \hspace{1ex}, 
		\label{main-text-gamma-tx-fv}
	\end{split}
\end{equation}

and

\begin{equation}
	\Gamma_{PL} = \frac{\Gamma_{C}^2 - \Gamma_{4} \Gamma_{TD} }{ \Gamma_3 - \Gamma_{TD} } \hspace{1ex},  
	\label{main-text-gamma-pl-fv}
\end{equation}

where 

\begin{eqnarray}
	\Gamma_{3} &\equiv&  \frac{\Gamma_{+} + \Gamma_{-}}{2} - \frac{\Gamma_{+} - \Gamma_{-}}{2} \sqrt{1 - 4 A^{+}_X A^{-}_X } \hspace*{1ex} , \nonumber \\
	\Gamma_{4} &\equiv&   \frac{\Gamma_{+} + \Gamma_{-}}{2} + \frac{\Gamma_{+} - \Gamma_{-}}{2} \sqrt{1 - 4 A^{+}_X A^{-}_X }  \hspace*{1ex} , \nonumber \\
	\Gamma_{C}^2 &\equiv& \frac{\left(\Gamma_{+} + \Gamma_{-} \right)^2 - \left(\Gamma_{+} - \Gamma_{-}\right)^2}{4} = \Gamma_{+} \Gamma_{-}  \hspace*{1ex} , 
	\label{Gamma-a-b-c}
\end{eqnarray}

are expressions that depend solely on experimentally obtainable values.

Equations (\ref{main-text-gamma-tx-fv}) - (\ref{Gamma-a-b-c}) make clear the dependence of an important physical parameter ($\Gamma_{TX}$) on experimentally accessible quantities (e.g. $\Gamma_{+}$, $\Gamma_{-}$, $A^{+}$ and $A^{-}$). However, these equations also reveal a challenge related to the dependence between the equations for the amplitudes ($A^{-}_X + A^{+}_X =1$). In general, we cannot solve equations (\ref{main-text-gamma-tx-fv}) and (\ref{main-text-gamma-pl-fv}) to obtain a value for $\Gamma_{TX}$ without either having a value for $\Gamma_{TD}$ or one more equation. For analogous reasons we cannot directly determine $\Gamma_{XD}$. However, the fact that all the involved rates must be positive can be used to constrain the possible values, as we will show in section \ref{Sect:application}.

Although in general the system of equation cannot be deterministically solved, the obstacle can be removed in the high temperature (HT) limit in which $\Gamma_{XT} \approx \Gamma_{TX}$  (see details in section 2.2 of the Supporting Information \cite{supp-mat}). This approximation essentially provides one additional equation relating the model rates, which leads to

\begin{equation}
	\Gamma_{PL} = \Gamma_{4} - \left( \Gamma_{+} - \Gamma_{-} \right) \sqrt{ A^{+}_X A^{-}_X }  \hspace{1ex}.  
	\label{high-t-MT}
\end{equation}

The usefulness of the model developed here is best appreciated when remembering that characteristic timescales for important physical processes cannot be directly accessed through measurements alone. The measured photoluminescence decay is the result of the interplay between all of the involved rates for competing photo-physical processes. Among these processes, perhaps the one easiest to misunderstand is $\tau_{PL}$. For example, in some works $\tau_{PL}$ is calculated by taking the average of the two measured decay times \cite{average-0,average-1}, but equation (\ref{main-text-gamma-pl-fv}) demonstrates that this is not correct. It is only by application of a model such as the one derived here that one can understand how to extract rates for the underlying physical processes from the experimental data.

\section{Synthesis and measurements of test samples}\label{Sect:Experimental}
We now introduce an experimental test system to which we will, in section \ref{Sect:application}, apply our model and analysis framework. The experimental test system consists of two samples (I and II), each of which is an ensemble of colloidally-synthesized semiconductor heterostructures composed of two quantum dots (QDs) separated by a nanorod as shown in figure \ref{figure-3}(a). Details on these structures and their intended optical functionality can be found in Refs.~\cite{ACSNANO-2018, TRPL-2021}. The important point for the analysis presented here is that the efficiency with which the target optical function can be implemented within such structures depends heavily on the rates of carrier transfer between different absorbing and emitting regions (e.g. the two QDs) and the rates of nonradiative decay \cite{Sellers2016a, Chen2018b, Chen2015b, Zhang2019}. There are two differences between samples I and II that influence these rates as described below. To assess how these differences in sample structure affect the rates of the underlying physical processes, we performed TRPL and PLQY studies of these nanostructures focused exclusively on emission from the lower-energy (smaller bandgap) QD on the left side of the structure as shown in figure \ref{figure-3}(a). As with other similar structures, the measured TRPL dynamics are biexponential \cite{Oron-biexponential,TRPL-2021}. Extracting, from this data, the underlying rates of carrier separation, carrier trapping, radiative recombination, and nonradiative recombination is typically challenging. Applying our model to this test case illustrates how it can be used to extract such information. We stress, however, that our model is not specific to these particular types of nanostructures nor to the particular optoelectronic device application for which they were designed.

\subsection{Sample synthesis}\label{sample}    
Both samples we consider here are dot/rod/dot heterostructures synthesized via a three step process that has been previously reported \cite{ACSNANO-2018,TRPL-2021}. In the first step, wurtzite CdSeTe core QDs are synthesized using a well-established hot-injection method \cite{Deutsch2012,Deutsch2013}. To synthesize the cores for Sample I (shown in orange in the top panel of figure \ref{figure-3}(a)), a solution with 90:10 ratio of Se:Te at room temperature is injected into a high temperature Cd-precursor solution of equal concentration. The reactivity difference between Se and Te causes a CdTe growth rate nearly twice that of CdSe, resulting in an inhomogeneous composition in the QD. The resulting QDs have a Te-rich center and a Se-rich outer layer, creating potential electron trap states due to the lower conduction band edge of CdSe as depicted in the top panel of figure \ref{figure-3}(a). To remove these trap states, sample II uses homogeneously alloyed CdSeTe cores synthesized under Cd-limited conditions (lower panel of figure \ref{figure-3}(a)) \cite{Bailey2003}.

For both samples, a nanorod (shown in blue in figure \ref{figure-3}(a)) is grown from the core CdSeTe QDs using a hot-injection seeded-growth method \cite{seededgrowth-1,seededgrowth-2}. For sample I, the rod consists only of CdS that is synthesized by injecting a mixture of S-precursor and the core CdSeTe QDs at room temperature into a hot Cd-precursor solution. CdS forms on the CdSeTe cores due to the lower activation energy for heterogeneous nucleation than for homogeneous nucleation. The morphology of the CdS is controlled by several factors. First, the (0001) facet of the wurtzite CdSeTe seeds is highly reactive, leading to preferential growth along the c-axis. Second, to further promote anisotropic nanorod growth, a mixture of short- and long-chain phosphonic acid ligands are used in the Cd-precursor solution \cite{seededgrowth-1,Lee2017}. Sample II contains alloyed CdSeS nanorods synthesized by adding a Se-precursor solution dropwise to the flask during the reaction. Adding increasing amounts of Se results in a gradient alloyed nanorod designed to funnel carriers away from the core QD and towards the right side of the structures depicted in figure \ref{figure-3}(a). 

For both samples, the final step of the synthesis is the growth of a CdSe QD on the end of the nanorod using successive ionic layer adsorption and reaction (SILAR) \cite{Deutsch2013}. Solutions of Cd and Se precursor are alternately added dropwise into a flask containing the core/rod particles. Growth of the CdSe emitter is monitored by photoluminescence spectroscopy during the reaction. When the emitter has reached the desired size, the Se-precursor is replaced with a S-precursor and a thin layer of CdS is grown to provide some surface passivation for the emitter QD. The resulting CdSe QDs are shown in green in figure \ref{figure-3}(a). 

Although the inner or surface character of the trapping defect states is unknown, the passivation offered by the CdS capping makes less likely that the involved traps are located on the core surfaces.

\subsection{Optical measurements}\label{experiment}

Detailed optical characterization methods for the study of these samples have been previously reported in \cite{ACSNANO-2018,TRPL-2021}. In brief, PLQY is measured using an integrating sphere (Labsphere) and calibrated with a rhodamine 101 standard (100\% QY). Samples are excited with 20 mW, 640 nm CW laser diode and the rhodamine standard is excited with 0.1 mW, 532 nm CW Verdi V10 diode laser. The emitted PL is collected in a circular fiber bundle from the integrating sphere exit port and directed to a Princeton Instruments Acton SpectraPro 2500i spectrometer. A 150 groove/mm grating directs the light toward a liquid-nitrogen-cooled charge-coupled device camera to measure the emission spectrum. The emission spectrum, absorbance factor, laser excitation power, and solvent refractive index for both the sample and rhodamine standard are inputs to a custom MATLAB program that calculates the PLQY ($\Phi$). 

\begin{figure}[H]
	\includegraphics[scale=0.5]{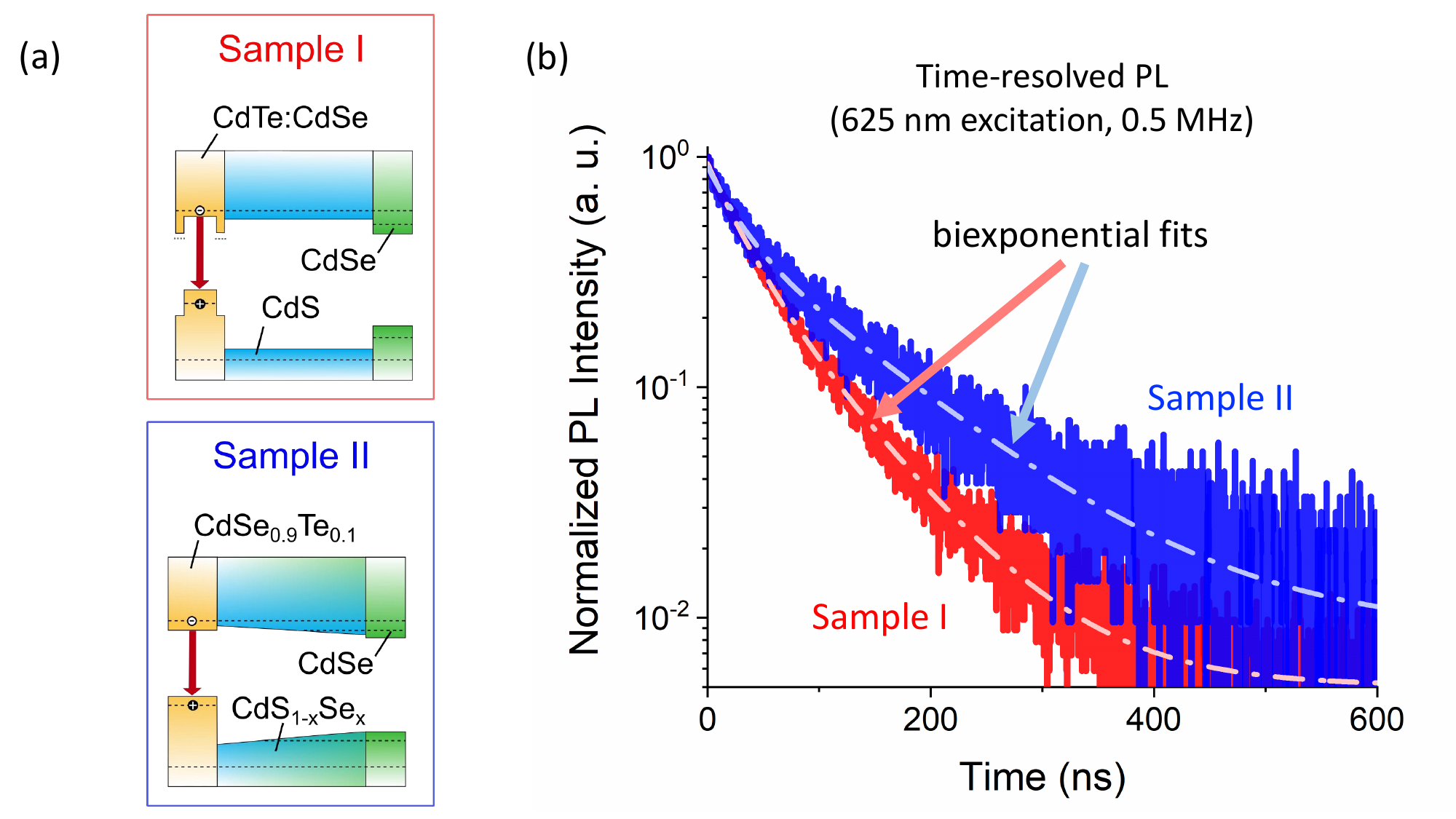}
	\caption{(a) Schematic band diagrams for the two core/rod/emitter heterostructures analyzed in this work.  Sample I consists of an inhomogenously alloyed CdSeTe core, a flat CdS rod, and a CdSe emitter.  Sample II consists of a homogeneously alloyed CdSeTe core, an alloyed CdSSe rod, and a CdSe emitter. (b) Time-resolved PL decay curves and corresponding 2-exponential fits for the two samples.}
	\label{figure-3}
\end{figure}

TRPL is measured using spectrally-resolved time-correlated single-photon counting (TCSPC). The data are shown in figure \ref{figure-3}(b). To obtain this data, the samples are excited by a 0.5 MHz tunable, pulsed laser emitting at 625 nm (Fianium WhiteLase). Importantly, photons with this excitation wavelength can only be absorbed in the CdSeTe QD because all other elements of the nanostructure have larger bandgaps. The PL emitted by recombination of carriers in the CdSeTe QD is detected by an avalanche photodiode (MPD Devices) that is correlated with the trigger signal from the laser using a PicoHarp 300 TCSPC system. While we have performed similar optical characterization experiments at other excitation and detection wavelengths, we focus here on TRPL and QY data obtained only for emission from the CdSeTe component of the dot/rod/dot heterostructures in order to demonstrate how the model presented here can be used to understand how the changes in structure between samples I and II influence the carrier dynamics (e.g. trapping and carrier escape rates) associated with this part of the nanostructure. The PL decay curves are fit using a biexponential function, $f(t) = A_X^-\exp\big({-\Gamma_- t}\big) + A_X^+\exp\big({-\Gamma_+ t}\big)$, where $\Gamma_-$ ($\Gamma_+$) is the inverse of the long (short) decay time constant and $A_X^-$ ($A_X^+$) is the amplitude coefficient for the long (short) decay component.  Table \ref{data-table} summarizes the relevant quantities obtained from these fits to the experimental data.  

\begin{table}[H]
	\centering
	\begin{center}
		\begin{tabular}{c|cc}
			& Sample I & Sample II \\
			\hline
			$\Gamma_+$ (ns$^{-1}$) & \hspace{1ex} 0.0302 $\pm$ 0.0005 \hspace{1ex} & \hspace{1ex} 0.0345 $\pm$ 0.0009 \\
			$\Gamma_-$ (ns$^{-1}$)	& \hspace{1ex} 0.0115 $\pm$ 0.0002 \hspace{1ex} & \hspace{1ex} 0.0083 $\pm$ 0.0001 \\
			$A^{+}_X$ (\%) & 0.66 $\pm$ 0.02  &  0.50 $\pm$ 0.01 \\
			$A^{-}_X$ (\%) & 0.34 $\pm$ 0.02  &  0.50  $\pm$ 0.01 \\
			$\Phi$ (\%)	& 66 $\pm$ 8 & 82 $\pm$ 7  
		\end{tabular}
		\caption{Values obtained from optical characterization of the test samples.}
		\label{data-table}
	\end{center}
\end{table}  


\section{Application of the model} \label{Sect:application}

For the sake of illustration, we now apply our model to the measurements of TRPL and PLQY performed on Samples I and II and reported in table \ref{data-table}. We first note that these data are obtained via optical measurements of ensembles of each sample type suspended in solution, which is the typical format for optical measurements of colloidally-synthesized nanostructures. The application of our model to ensemble measurements raises two important questions that originate in the fact that ensembles of nanoparticles are in general inhomogeneous due to differences in size, shape, composition, or defects. Among these, size non-uniformity is particularly relevant and is primarily responsible for the broadening of PL from ensembles of emitters synthesized in a single batch \cite{size-inhomogeneity-1,size-inhomogeneity-2,size-inhomogeneity-3}. The first question that should be asked before applying a model such as the one developed here is whether or not the measured biexponential decay may originate primarily from a bimodal distribution in the sizes of the particles. We have verified, via TEM imaging, that this is not the case for the samples we consider here \cite{ACSNANO-2018,TRPL-2021}. Moreover, the experimental data reported here was all obtained using laser excitation conditions that prevent emission from the CdSe QD. In other words, we have verified that we are in situation (ii) as defined in section \ref{Sect:Intro}. The second question that should be asked is whether a single particle model can be applied to an ensemble that has a statistical variation in size, shape, and composition. This is normally the case in most experiments and is the case for the test data we consider here. In section 3 of the Supporting Information \cite{supp-mat}, we show that the application of this model to ensemble measurements is valid and results in the extraction of representative values for the average member of the ensemble, provided the sample has a unimodal distribution as is the case here.

It is worth mentioning that besides variations in the structural features of the single emitter, a diversity in number and type of trapping defects can be expected among the emitters constituting the ensemble. It is indeed a limitation of any single particle-based model when applied to ensembles. In applying the developed model to these collective samples, we are assuming that their biexponential decay reflects a prevalence of the one-trap-per-emitter component in the distribution for number of traps. This, in consistence with the obtained result in equation (\ref{solutions}), yielded by considering a single trapping state per emitter.          

In applying the model to these samples, we will ignore the exciton fine structure because the data were obtained at room temperature. At low temperature, photon emission from dark states usually prevails in the PL signal because thermal repopulation of the bright states after quick spin relaxation is inefficient \cite{quick-spin-relaxation,Efros-dark-exciton}. However, as temperature increases, the phonon-assisted population transfer between bright and dark exciton states occurs at a time scale that is much faster than the observed dynamics in our samples. Thus, in this case it is valid to assume emission only from the neutral bright exciton state, since emission from the exciton dark states is much weaker and slower \cite{Zunger-dark-exciton,Nanoscale-dark-exciton}. 

Direct insertion of the values from table \ref{data-table} into equation ({\ref{cleared-Gxg}}) gives values for the radiative lifetime $\tau_{rad}=\frac{1}{\Gamma_{XG}}$ of 78.13 ns for sample I and 90.91 ns for sample II. These radiative lifetimes are approximately 3x longer than those reported in the literature for CdSe nanocrystals \cite{almost-unit-QY, multiexp-2}. We suspect this is related to the reduced electron-hole overlap in our samples, in which the embedding of the CdSeTe quantum dots in CdS or CdSSe nanorods reduces the electron confinement in comparison to core-only or spherical core-shell quantum dots.      

We next analyze the system of equations to determine the range of possible values for several parameters. Recalling that $ \Gamma_{PL} \equiv \Gamma_{XG} + \Gamma_{XD} $ and $ \Gamma_{4} = \Gamma_{PL} + \Gamma_{XT} $, it can be figured out that $\Gamma_{XG} \le \Gamma_{PL} \le \Gamma_4 $. Furthermore, from equation (\ref{main-text-gamma-pl-fv}) it can be seen that the function $\Gamma_{PL}(\Gamma_{TD})$ in the domain $\Gamma_{TD}\ge0$ starts with a positive value $\Gamma_{PL} (0) = \frac{\Gamma_{C}^2}{\Gamma_3} $ and becomes 0 at $\Gamma_{TD} = \frac{\Gamma_{C}^2}{\Gamma_4}$ (i.e. $\Gamma_{PL} (\frac{\Gamma_{C}^2}{\Gamma_4}) = 0$). It is worth noting that $\Gamma_{PL} (\Gamma_{TD})$ is monotonically decreasing in the interval ($-\infty,\frac{\Gamma_{C}^2}{\Gamma_4}$) and continues asymptotically diverging towards $-\infty$ at $\Gamma_{TD}=\Gamma_3$ (i.e. $\Gamma_{PL} (\Gamma_3)= -\infty$). Thus, the maximum possible value for $\Gamma_{PL}$ is $\frac{\Gamma_{C}^2}{\Gamma_3}$, satisfying $\frac{\Gamma_{C}^2}{\Gamma_3} \le \Gamma_{4}$. Using these constraints we define $\Gamma_{PL}^{Min} \equiv \Gamma_{XG}$ and $\Gamma_{PL}^{Max} \equiv \frac{\Gamma_{C}^2}{\Gamma_3}$, which limit the range of possible values for $\Gamma_{PL}$. Given the definition of $\Gamma_{4}$, this also implies that $ \Gamma_{XT}^{Min} \equiv \Gamma_{4} - \frac{\Gamma_{C}^2}{\Gamma_3} $ and $\Gamma_{XT}^{Max} \equiv \Gamma_{4} - \Gamma_{XG}$.

We next invert equation (\ref{main-text-gamma-pl-fv}) with respect to $\Gamma_{TD}$ to obtain

\begin{equation}
	\Gamma_{TD} = \frac{\Gamma_{C}^2 - \Gamma_{3} \Gamma_{PL} }{ \Gamma_4 - \Gamma_{PL} } \hspace{1ex},  
	\label{main-text-gamma-td}
\end{equation}

from which we can know the corresponding $\Gamma_{TD}$ for a given $\Gamma_{PL}$. Because the function $\Gamma_{TD} (\Gamma_{PL})$ (equation (\ref{main-text-gamma-td})) is monotonically decreasing in the interval $(0,\Gamma_4)$, the maximum possible value for $\Gamma_{TD}$ is that for which $\Gamma_{PL}$ is minimum. Hence, because $\Gamma_{TD} \ge 0$ it can be stated that $\Gamma_{TD}^{Min} \equiv 0$ and $\Gamma_{TD}^{Max} \equiv \frac{\Gamma_{C}^2 - \Gamma_{3} \Gamma_{PL}^{Min} }{ \Gamma_4 - \Gamma_{PL}^{Min} } = \frac{\Gamma_{C}^2 - \Gamma_{3} \Gamma_{XG} }{ \Gamma_4 - \Gamma_{XG}}$. Consequently, $\Gamma_{TX}^{Min} \equiv \Gamma_{3} - \Gamma_{TD}^{Max}$ and $\Gamma_{TX}^{Max} \equiv \Gamma_{3} $.        

\begin{figure}[H]
	\begin{center}
		\includegraphics[scale=0.7]{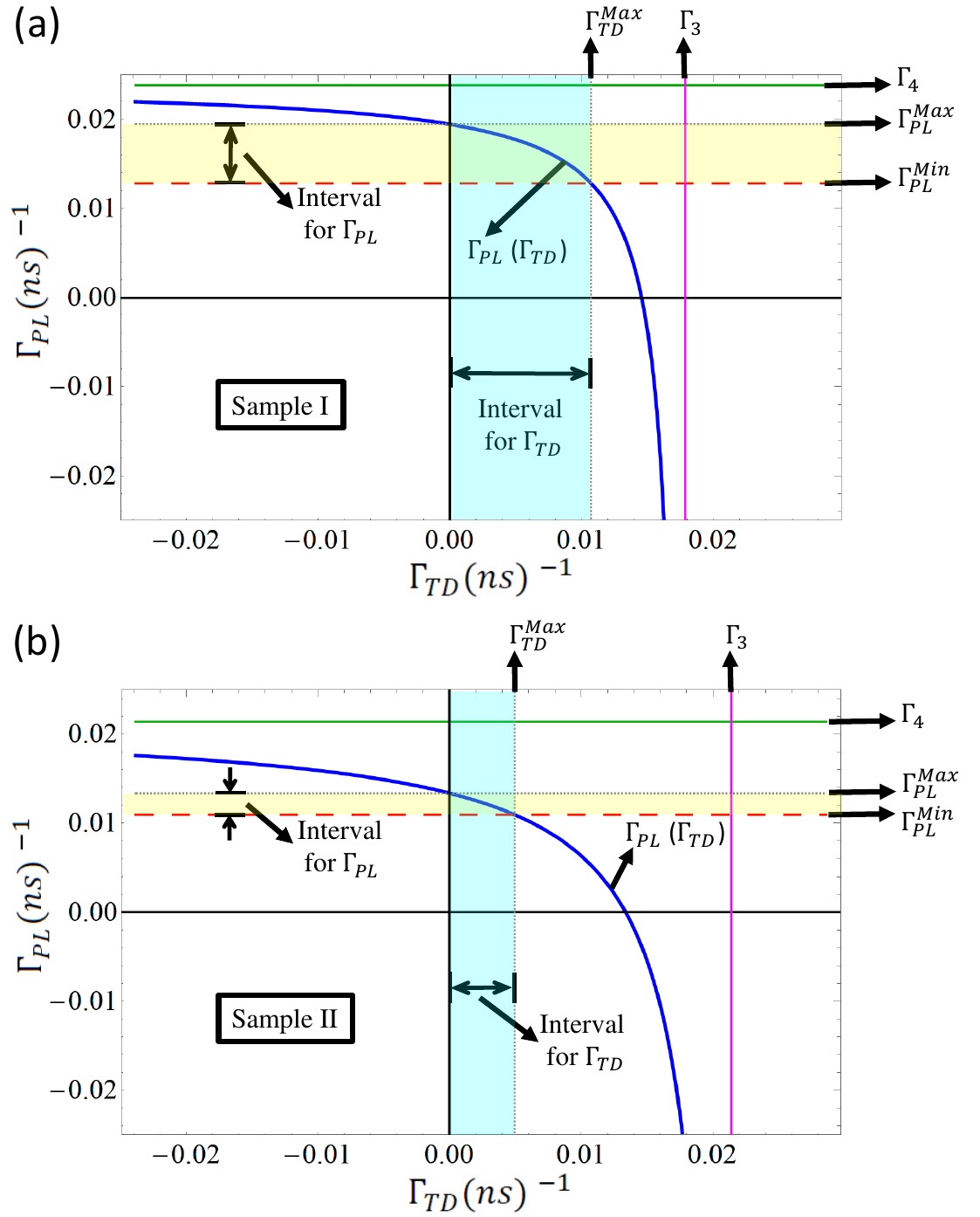}
		\caption{(a) $\Gamma_{PL} (\Gamma_{TD})$ for the experimentally studied sample I, where the allowed intervals for $\Gamma_{PL}$ and $\Gamma_{TD}$ are highlighted by yellow and blue fringes, respectively. (b) As in (a) but for the studied sample II.}
		\label{image-4}
	\end{center}
\end{figure}

We now have limits on the range of possible values for $\Gamma_{PL}$, $\Gamma_{TD}$, $ \Gamma_{XT}$, and $\Gamma_{TX}$. Figure \ref{image-4} depicts these limits graphically. In figure \ref{image-4}(a) we show with the solid blue line the function $\Gamma_{PL} (\Gamma_{TD})$ calculated using equation (\ref{main-text-gamma-pl-fv}) and the values of $\Gamma^2_{C}$, $\Gamma_{4}$, and $\Gamma_{3}$ computed using equations (\ref{Gamma-a-b-c}), from the experimentally measured values for sample I as reported in table \ref{data-table}. The range of values for $\Gamma_{PL}$ that are possible given the constraints described above are highlighted by the yellow rectangle. The range of values for $\Gamma_{TD}$ that are possible given the constraints described above are highlighted by the blue rectangle. The intersection of these rectangles indicates the range of values for $\Gamma_{PL}$ and $\Gamma_{TD}$ that are compatible with both the experimental data and the physical constraints. Figure \ref{image-4}(b) shows the analogous analysis for sample II. 

\begin{table}[t]
	\centering
	\begin{center}
		\begin{tabular}{c|cc}
			& Sample I & Sample II \\
			\hline
			$\Gamma_{XG}$  & \hspace{1ex} [0.0112,0.0144] \hspace{1ex} & \hspace{1ex} [0.0100,0.0120] \\
			$\Gamma_{XD}$ 	& \hspace{1ex} [0,0.0066] \hspace{1ex} & \hspace{1ex} [0,0.0024] \\
			$\Gamma_{PL}$  & \hspace{1ex} [0.0128,0.0194] \hspace{1ex} & \hspace{1ex} [0.0110,0.0134] \\
			$\Gamma_{XT}$  & \hspace{1ex} [0.0044,0.0110] \hspace{1ex} & \hspace{1ex} [0.0080,0.0104] \\
			$\Gamma_{TX}$  & \hspace{1ex} [0.0071,0.0179] \hspace{1ex} & \hspace{1ex} [0.0165,0.0214] \\
			$\Gamma_{TD}$	& \hspace{1ex} [0,0.0107] \hspace{1ex} & \hspace{1ex} [0,0.0049]  
		\end{tabular}
		\caption{Plausible intervals for the relevant rates in ns$^{-1}$, calculated by application of the model to the experimental values obtained from the studied samples.}
		\label{results-table}
	\end{center}
\end{table}  

Table \ref{results-table} summarizes the range of possible values for the model rates $\Gamma_{XG}$, $\Gamma_{XD}$, $\Gamma_{PL}$, $\Gamma_{XT}$, $\Gamma_{TX}$ and $\Gamma_{TD}$ determined by applying the constraints to the parameters calculated from the measured samples. Because the extremes of these intervals are directly computed from values extracted from experimental measurements (table \ref{data-table}), the error bars on the corresponding experimental measurements directly translate into confidence intervals on the bounds of the allowable ranges. Carrying out the respective error propagation analysis (see section 4 of the Supporting Information \cite{supp-mat}), we find that the uncertainties in the interval extremes range from 3\% to 13\%. e.g. $\Gamma_{PL}^{Min} = 0.0128 \pm 0.0016 $ and $\Gamma_{PL}^{Max} = 0.0193 \pm 0.0011$ ($\Gamma_{PL}^{Min} = 0.0110 \pm 0.0010$ and $\Gamma_{PL}^{Max} = 0.0133 \pm 0.0006 $) for sample I (II). The intervals for $\Gamma_{XG}$, whose midpoints are the single values obtained for each sample from equation (\ref{cleared-Gxg}), are calculated using the same error analysis.     

While it would be much more desirable to have defined values for the rates with narrow confidence intervals, the information in table \ref{results-table} is still valuable. For instance, comparison of $\Gamma_{XG}$ and $\Gamma_{XD}$ for samples I and II reveals that the direct radiative rates ($\Gamma_{XG}$) are similar in both cases, but the nonradiative rate ($\Gamma_{XD}$) is larger in sample I than in sample II. This is consistent with the higher PLQY measured for sample II. Similar behavior is observed for $\Gamma_{TD}$, which is also a nonradiative rate and appears larger in sample I than in sample II. That $\Gamma_{XG}$ should be similar in the two samples is reasonable because both emitter cores are nanocrystals of similar size and composition (CdSe$_{0.9}$Te$_{0.1}$). 

It is also interesting to consider the order of magnitude of the rates for trapping and the release from traps ($\Gamma_{XT}$ and $\Gamma_{TX}$). The obtained intervals for the trapping rates are roughly similar in both samples, but the rates for the release from traps differ more noticeably. Furthermore, the corresponding intervals overlap for sample I but do not for sample II. This suggests that $\Gamma_{TX}$ is somehow more sensitive to the band structure details than $\Gamma_{XT}$. This issue will be addressed again ahead. 

For the sake of intuitive comprehension of the results, table \ref{times-v2} shows the same intervals as in table \ref{results-table}, but in terms of times (inverses of the corresponding rates). Additionally, a representative value for each interval is also shown. In all cases the representative lifetime is computed as the inverse of a rate lying within the calculated allowed interval. In most cases, the chosen rate is calculated as the inverse of the midpoint of the interval. For intervals that include 0, the chosen rate is the inverse of the maximum value. These numbers provide an estimate of the time scale for each of the transition processes.

\begin{table}[H]
	\centering
	\begin{center}
		\begin{tabular}{c|ccccc}
			\hspace{1ex} & \hspace{5ex} Sample &  I \hspace{5ex} & \hspace{1ex} & 
			\hspace{5ex} Sample &  II \hspace{5ex} \\
			\hspace{1ex} & \hspace{1ex} Interval \hspace{1ex} & \hspace{1ex} Rep. Value \hspace{1ex} & \hspace{3ex} & \hspace{1ex} Interval \hspace{1ex} & \hspace{1ex} Rep. Value  \hspace{1ex} \\
			\hline
			$\tau_{rad}$ (ns) & [69.44,89.29] & \hspace{1ex}  78.13 \hspace{1ex} & 
			\hspace{3ex} & [83.33,100] & \hspace{1ex}  90.91 \\
			$\tau_{XD}$ (ns) & [151.52,$\infty$] & \hspace{1ex} $\ge$ 151.52 \hspace{1ex} & \hspace{3ex} & [416.67,$\infty$] & \hspace{1ex} $\ge$ 416.67\\ 
			$\tau_{PL}$ (ns) & [51.55,78.13] & \hspace{1ex} 62.11 \hspace{1ex} & 
			\hspace{3ex} & [74.66,90.91] & \hspace{1ex} 81.97 \\
			$\tau_{XT}$ (ns) & [90.91,227.27] & \hspace{1ex} 129.87 \hspace{1ex} & \hspace{3ex} & [96.15,125] & \hspace{1ex} 108.7 \\  
			
			$\tau_{TX}$ (ns) & [55.87,140.85] & \hspace{1ex} 80  \hspace{1ex} & \hspace{3ex} & [46.73,60.61] & \hspace{1ex} 52.77 \\
			$\tau_{TD}$	(ns) & [93.46,$\infty$] & \hspace{1ex} $\ge$ 93.46 
			\hspace{1ex} & \hspace{3ex} & [204.08,$\infty$] & \hspace{1ex} $\ge$ 204.08  
		\end{tabular}
		\caption{Time intervals and representative values associated to the plausible intervals shown in table \ref{results-table} ($\tau_{i} = 1/\Gamma_{i}$,  $\tau_{rad} \equiv \tau_{XG}$). For the rates $\Gamma_{XG}$, $\Gamma_{PL}$, $\Gamma_{XT}$ and $\Gamma_{TX}$ the representative values are the inverses of the midpoints of the plausible intervals, while for  $\Gamma_{XD}$ and $\Gamma_{TD}$ the representative values are the inverses of the interval maxima.  }
		\label{times-v2}
	\end{center}
\end{table}

It is important to note that these numbers must be considered with care because there is no basis by which we can conclude that any particular value within the allowed ranges is more likely than any other. However, considering numbers that lie within the allowed ranges already sheds light on the trapping and detrapping processes. For example, in samples in which these processes are not particularly long compared to the conventional $\tau_{XG}$ (radiative) and  $\tau_{XD}$ (nonradiative) lifetimes, the monoexponential limit is far from suitable for describing the time dependent PL, as can be seen in equations (\ref{rates-2}) and (\ref{amplitudes-X}). Indeed, in our studied sample II, in which the biexponential feature is so obvious that the two observed PL decay time constants exhibit the same amplitude coefficient, the nonradiative process associated to the time $\tau_{XD}$ is found to be significantly larger than $\tau_{XT}$ and $\tau_{TX}$ ($\Gamma_{XD} \ll \Gamma_{TX}$). 

If the high temperature limit is assumed, the likelihood intervals become specific values. Table \ref{results-table-HT} shows those values for the different transition rates that can be computed within the HT approximation for the two studied samples. 

\begin{table}[H]
	\centering
	\begin{center}
		\begin{tabular}{c|cccc}
			& \hspace{1ex} & Sample I & \hspace{1ex} & Sample II \\
			\hline
			$\Gamma_{XG}$ \hspace{1ex} ($\tau_{rad}$)  & \hspace{1ex} & \hspace{1ex} 0.0128 \hspace{1ex} (77.91) \hspace{1ex} & \hspace{1ex} & \hspace{1ex} \dotuline{0.0109} \hspace{1ex} (\dotuline{91.22}) \\
			$\Gamma_{XD}$ \hspace{1ex} ($\tau_{XD}$)	& \hspace{1ex} & \hspace{1ex} 0.0021 \hspace{1ex} (465.74) \hspace{1ex} & \hspace{1ex} & \hspace{1ex} -\dotuline{0.0026} \hspace{1ex} (-\dotuline{374.44}) \\
			$\Gamma_{PL}$ \hspace{1ex} ($\tau_{PL}$) & \hspace{1ex} & \hspace{1ex} 0.0149 \hspace{1ex} (66.75) \hspace{1ex} & \hspace{1ex} & \hspace{1ex} \dotuline{0.0082} \hspace{1ex} (\dotuline{120.6}) \\
			$\Gamma_{XT}$ \hspace{1ex} ($\tau_{XT}$) & \hspace{1ex} & \hspace{1ex} 0.0089 \hspace{1ex} (112.78) \hspace{1ex} & \hspace{1ex} & \hspace{1ex} \dotuline{0.0131} \hspace{1ex} (\dotuline{76.36}) \\
			$\Gamma_{TX}$ \hspace{1ex} ($\tau_{TX}$) & \hspace{1ex} & \hspace{1ex} 0.0089 \hspace{1ex} (112.78) \hspace{1ex} & \hspace{1ex} & \hspace{1ex} \dotuline{0.0131} \hspace{1ex} (\dotuline{76.36}) \\
			$\Gamma_{TD}$ \hspace{1ex} ($\tau_{TD}$)	& \hspace{1ex} & \hspace{1ex} 0.0090 \hspace{1ex}(111.21)\hspace{1ex} & \hspace{1ex} & \hspace{1ex} \dotuline{0.0083} \hspace{1ex} (\dotuline{120.6}) 
		\end{tabular}
		\caption{Relevant rates (times) in ns$^{-1}$ (ns) within the HT approximation, calculated by application of the model to the experimental values obtained from the studied samples. The values for sample II are shown for the sake of completeness, but the dotted underlines are intended to emphasize that these values are not reliable because the trap energies for sample II do not satisfy the HT approximation at room temperature.}
		\label{results-table-HT}
	\end{center}
\end{table}

The first important conclusion we can draw from the values presented in table \ref{results-table-HT} is that the fact that one of the rates is found to be negative for sample II clearly indicates that something went wrong in applying the HT approximation to sample II. This should have been expected from the beginning because it was already noted that the calculated intervals for $\Gamma_{XT}$ and $\Gamma_{TX}$ did not overlap for that sample (see table \ref{results-table}). Thus, obviously $\Gamma_{XT} \approx \Gamma_{TX}$ can not be true. This leads to two points: a) one must use caution when applying the high temperature approximation instead of simply taking it for granted, as has been done in some previously published work \cite{PhysRevLett.high-T}; b) the fact that the calculated intervals for $\Gamma_{XT}$ and $\Gamma_{TX}$ do not overlap for sample II tells us that the energy difference $| E_X - E_T |$ for sample II is of the order of $k_B T$ at room temperature \textbf{and} significantly larger than $| E_X - E_T |$ for sample I, where the allowed intervals for $\Gamma_{XT}$ and $\Gamma_{TX}$ do overlap at the same temperature. This provides important information about the relative depth of trap states in sample II, which may be correlated to synthesis conditions. Furthermore, for sample II the fact that $\Gamma_{XT} < \Gamma_{TX}$, tells us that $ E_X - E_T < 0 $ (see subsection 2.2 of the Supporting Information \cite{supp-mat}).  

The second important conclusion comes from focusing on sample I, where the high temperature approximation appears to be suitable. We observe that the computed values presented in table \ref{results-table-HT} are quite similar to the midpoints  of the corresponding intervals that are reported in table (\ref{times-v2}). In particular, we note that the PL and trapping times are found to be $\tau_{PL}=66.75$ ns and $\tau_{XT}=112.78$ ns, respectively. We conclude this section with two comments about these two values that illustrate the importance of this model.

First, the value we obtain for the PL decay time shows that the usual procedure of taking an amplitude-weighted average time as the decay constant for systems exhibiting multiexponential decay \cite{average-0,average-1,average-2,average-3} does \textbf{not} yield an accurate number for $\tau_{PL}$ in single-species ensembles. If the amplitude-weighted average method were used to estimate the PL decay time for sample I, the obtained value would be $\frac{A^{+}_X}{\Gamma_{+}} + \frac{A^{-}_X}{\Gamma_{-}} \equiv \tau_{PL}^{'} = 51.42$ ns, which represents an underestimation by more than 20\% relative to the more precise value extracted using the methods reported here. This could, in turn, induce underestimation of the radiative lifetime if the expression $\tau_{rad}^{'} = \frac{\tau_{PL}}{QY}$ is used. We suspect this may be the case in prior reports; for example Ref.~\cite{multiexp-2} reports radiative decay times ranging from 26.5 to 66.3 ns, but application of our model yields $\tau_{rad} = 77.92$ ns.

Second, the value we obtain for the trapping time ($\tau_{XT}$) for sample I is around one hundred nanoseconds, only a factor of two larger than that reported in Ref.~\cite{multiexp-2}. However, Ref.~\cite{multiexp-2} made the assumption that $\Gamma_{TD}=0$. If we applied such an assumption in our model, we would obtain $\tau_{XT}^{'} = 60$ ns, a factor of two smaller than what we obtain. In other words, the results we obtain with our model are consistent with prior models when the same assumptions are employed. However, the model we present provides a framework for obtaining values for the rates of important physical processes \textit{without} the restrictions imposed by making these assumptions in all cases.

The application of the model shown in this section is just an example of how the developed framework can be implemented for analyzing data from time-resolved optical measurements exhibiting biexponential decays. It may be suitable for obtaining valuable information on the relevant transition rates from experiments on ensembles or, even more, on single emitters. Some other scenarios in which it could be applied include low-temperature measurements for studying phonon-assisted dark-bright exciton dynamics in zero-, one- and two-dimensional semiconductors.     

\section{Conclusions}\label{Sect:Conclusion}

We developed a model for analyzing biexponential photoluminescence decay data to extract rates for physical processes including radiative recombination, nonradiative relaxation, trapping, and detrapping. This model led to equations that directly connect the rates of these different physical processes to experimentally accessible quantities like the two photoluminescence decay times, amplitude coefficient for each, and photoluminescence quantum yield. We applied this model to values extracted from optical characterization of two rod-embedded double-quantum-dot heterostructures designed to change carrier escape and trapping rates. Application of the model to these samples, without approximations, yields limits on the range of possible values for the rate of each physical process. These limits provide useful information on the time scales of relevant dynamical processes. In particular, the trapping and detrapping times were found to be of order one hundred nanoseconds. We next showed how the interval analysis can be used to determine when samples are in the limit at which temperatures are sufficiently high to justify the assumption that the trapping and detrapping rates are equal. When samples are in this limit, as one of ours is, the model allows for determination of specific values for all of transition rates. A comparison of the results calculated with this model to those calculated using less general models from the literature reveal that the limited models underestimate, by more than 20\%, the values of parameters such as the radiative lifetime. The model and analysis framework developed here can be applied  to improve characterization and understanding of the optical properties of a wide range of nanostructured semiconductor emitters for optoelectronic device applications.

\section{Acknowledgments}
	
	M. F. D, T. A. W. and J. M. C. acknowledge partial support from the Delaware Energy Institute and partial support from the National Science Foundation through the University of Delaware Materials Research Science and Engineering Center (MRSEC, DMR-2011824), including partial support of this project through the MRSEC-affiliated Partnership for Research and Education (PREM) in Soft Matter Research \& Technology and Quantum Confinement Materials Design (SMaRT QD, DMR-2122158). T. A. W. and J. M. C. acknowledge support from the Delaware Space Grant College and Fellowship program 	(NASA Grant 80NSSC20M0045). 
	
	H.Y.R. acknowledges support from the Colombian SGR through project BPIN2021000100191 and from the Research Division of UPTC through Project No. SGI-3378.
	

\section{Supporting Information}
	
	The Supporting Information is available free of charge at (https://... to be provided) 
	
	It contains four sections: S.1 Amplitudes, ND limit and PLQY. S.2 Equation Reduction. S.3 Fitting Ensemble Measurements with a Single Particle Model and S.4 Error propagation.
	
	

\bibliographystyle{apsrev4-2}
\bibliography{Re-resubmission-framework-bibliography}

\end{document}